\documentclass[11pt,a4paper]{article}
\pdfoutput=1
\usepackage{jcappub}

\usepackage[T1]{fontenc}
\usepackage{graphicx}
\usepackage{amsfonts,amsmath,amssymb,tensor}
\usepackage{mathtools}
\usepackage{braket}
\usepackage{bm}
\usepackage{enumerate}
\usepackage{ulem}
\usepackage{comment}
\usepackage{tikz}
\usetikzlibrary{shapes, calc}
\usepackage[dvipsnames]{xcolor}

\begin{document}

\title{Reviving the Affleck-Dine Curvaton Scenario with Compensated Isocurvature Perturbations}
\author[a,b,c]{Masahiro Kawasaki,}
\author[b]{Shunsuke Neda}

\affiliation[a]{Kavli IPMU (WPI), UTIAS, University of Tokyo, Kashiwa, 277-8583, Japan}\affiliation[b]{ICRR, University of Tokyo, Kashiwa, 277-8582, Japan}
\affiliation[c]{Physics Division, Faculty of Science, Kanagawa University, Kanagawa 221-8686, Japan}

\abstract{%
We revisit the Affleck-Dine curvaton scenario in the framework of supersymmetry, taking Q-ball formation into account. 
In this scenario, a scalar field carrying baryon number, namely the Affleck-Dine field, generates both the curvature perturbations and the baryon asymmetry of the Universe. 
However, correlated baryon isocurvature perturbations are inevitably generated, resulting in excessively large correlated matter isocurvature perturbations. 
We propose a mechanism to suppress them, in which the correlated baryon isocurvature perturbations are canceled by anti-correlated dark matter isocurvature perturbations, thereby realizing compensated isocurvature perturbations. 
We show that the scenario can simultaneously account for the baryon asymmetry and the observed curvature perturbations of the Universe without generating excessively large matter isocurvature perturbations.
}


\emailAdd{masahiro.kawasaki@ipmu.jp}
\emailAdd{neda@icrr.u-tokyo.ac.jp}

\maketitle
\section{Introduction}
\label{sec: intro}

Asymmetry between matter (baryons) and antimatter (antibaryons) is one of the outstanding problems in modern cosmology. 
The baryon asymmetry is known to have existed by the epoch of Big Bang nucleosynthesis (BBN), with the baryon-to-entropy ratio estimated to be $\eta_B\sim10^{-10}$. 
Moreover, any pre-existing baryon asymmetry would have been exponentially diluted by inflation. 
Therefore, the baryon asymmetry of the Universe must have been generated after inflation and before BBN.
The Affleck-Dine mechanism~\cite{Affleck:1984fy} is one of the most promising mechanisms for generating the baryon asymmetry (baryogenesis). 
In this mechanism, a complex scalar field carrying baryon number acquires a large field value during inflation and begins to oscillate after inflation. 
At the onset of oscillation, the scalar field is kicked in the phase direction by a phase-dependent potential and begins to rotate in the complex field plane. 
This rotational motion generates a net baryon number.

The Affleck-Dine mechanism is most naturally realized in the framework of supersymmetry (SUSY)~\cite{Dine:1995kz}. 
In the minimal supersymmetric standard model (MSSM), the scalar potential possesses many flat directions. 
Since these flat directions are composed of squarks, sleptons, and Higgs fields, they generally carry baryon or lepton number.
Let us denote one of these flat directions by the Affleck-Dine (AD) field and assume that it carries baryon number. 
The dynamical evolution of the AD field then generates the baryon asymmetry, as described above.

During inflation, the AD field acquires a large field value and quantum fluctuations of amplitude $H_I/(2\pi)$, where $H_I$ is the Hubble parameter during inflation, provided that its potential is sufficiently flat. 
After inflation, the oscillating AD field can make a significant contribution to the energy density of the Universe before eventually decaying into radiation. 
Through this process, the fluctuations of the AD field generate curvature perturbations. 
Thus, the AD field can play the role of a curvaton~\cite{Moroi:2001ct,Enqvist:2001zp,Lyth:2001nq}.

The possibility that the AD field acts as a curvaton was first discussed in Ref.~\cite{Hamaguchi:2003dc}, where it was pointed out that this scenario produces excessively large baryon isocurvature perturbations. 
In particular, fluctuations of the radial component of the AD field inevitably induce isocurvature perturbations correlated with the curvature perturbations. 
Such correlated isocurvature perturbations are stringently constrained by CMB observations~\cite{Planck:2018jri}, which makes the AD curvaton scenario unviable.
More recently, Ref.~\cite{Ghoshal:2024wom} studied the AD curvaton scenario in a non-SUSY framework and showed that the isocurvature perturbations can be suppressed by introducing a quadratic phase-dependent potential. 
In SUSY, however, the phase-dependent potential arises from a non-renormalizable operator, and the resulting isocurvature perturbations cannot be similarly suppressed.

In this paper, we revisit the AD curvaton scenario in the framework of SUSY and propose a different mechanism for suppressing the correlated isocurvature perturbations. 
In our scenario, the correlated baryon isocurvature perturbations are canceled by anti-correlated dark matter isocurvature perturbations, thereby realizing \textit{compensated isocurvature perturbations}~\cite{Holder:2009gd,Gordon:2009wx,Grin:2011tf,Grin:2013uya,Harigaya:2014bsa} and evading the CMB constraints. 
Furthermore, we take into account Q-ball formation~\cite{Coleman:1985ki}, which generically accompanies the Affleck-Dine mechanism and plays an essential role in the cosmological evolution of the AD field~\cite{Dvali:1997qv,Enqvist:1997si,Kasuya:1999wu}. 
We find that the SUSY AD curvaton scenario can simultaneously generate the baryon asymmetry and the observed curvature perturbations of the Universe without producing excessively large isocurvature perturbations.

The rest of this paper is organized as follows.
In Sec.~\ref{sec:curvaton}, we describe the AD curvaton scenario in the framework of SUSY and discuss the generation of the baryon asymmetry, Q-ball formation, and curvature perturbations.
In Sec.~\ref{sec:iso_curv_perterbation}, we review the isocurvature perturbation problem and present our solution based on compensated isocurvature perturbations.
In Sec.~\ref{sec:successful_scenario}, we derive the conditions for a successful AD curvaton scenario and determine the allowed parameter space.
Finally, Sec.~\ref{sec:summary} is devoted to the summary and discussion.

\section{Affleck-Dine curvaton scenario}
\label{sec:curvaton}

We consider the AD field as a curvaton.
The AD curvaton scenario proceeds as follows:
\begin{enumerate}
    \item The AD field acquires fluctuations during inflation.
    \item The AD field begins to oscillate and generates the baryon asymmetry through the Affleck-Dine mechanism.
    \item Q-balls are formed.
    \item The Q-ball energy density fraction increases after reheating.
    \item The Q-balls decay into radiation, thereby generating curvature perturbations.
\end{enumerate}
In the following, we describe each step of the scenario.
For simplicity, we assume throughout this paper that the contribution of the inflaton to the curvature perturbations is negligible.

\subsection{Affleck-Dine field during inflation}
\label{sec:during_inflation}

The potential of the AD field $\Phi$ is lifted by SUSY-breaking effects and non-renormalizable operators and is given by
\begin{equation}
    \label{eq:AD_potential}
   V(\Phi)=m_\Phi^2|\Phi|^2
   +cH^2|\Phi|^2
   +\lambda^2\frac{|\Phi|^{2(n-1)}}{M_{\mathrm{pl}}^{\,2(n-3)}}
   +\left(
   \frac{a_M \lambda m_{3/2}}{M_{\mathrm{pl}}^{\,n-3}}\Phi^n
   +\mathrm{h.c.}\right),
\end{equation}
where $m_\Phi$ is the soft SUSY-breaking mass, $H$ is the Hubble parameter, and the second term represents the Hubble-induced mass term. 
The third term is a non-renormalizable operator characterized by an integer $n\geq4$, while the last term is the so-called $A$-term. 
Here, $M_{\mathrm{pl}}$ denotes the reduced Planck mass, $m_{3/2}$ is the gravitino mass, and $c$, $\lambda$, and $a_M$ are dimensionless constants.

In general, during inflation, when $H\gg m_\Phi$, the Hubble-induced mass term and the non-renormalizable term dominate the potential. 
In the present scenario, however, we assume that both contributions are negligible, namely, $|c|\ll1$ and $|\lambda|\ll1$. 
The AD field can therefore be treated as an approximately massless field and remains nearly constant during inflation.

The AD field is a complex scalar field, which can be parameterized as
\(
\Phi=\phi e^{i\theta}/\sqrt{2}.
\)
During inflation, it acquires quantum fluctuations given by
\begin{align}
\label{eq:AD_fluctuation}
  \delta\phi_i
  =
  \frac{H_I}{2\pi},\qquad
  \delta\theta_i
  =
  \frac{H_I}{2\pi\phi_i}.
\end{align}
Here $\phi$ and $\theta$ denote the radial and angular components of the AD field, respectively, the subscript $i$ indicates the values at the end of inflation, and $H_I$ is the Hubble parameter during inflation.

\subsection{Affleck-Dine field oscillation}
\label{sec:AD_oscillationn}

The AD field begins to oscillate when the Hubble parameter becomes comparable to the soft SUSY-breaking mass, i.e.,
$H_{\mathrm{osc}}\simeq m_\Phi$.
Here and in what follows, the subscript ``osc'' denotes quantities evaluated at the onset of oscillation.
In our scenario, we assume that
\begin{align}
    \frac{\lambda^2}{M_{\mathrm{pl}}^{2n-6}}
    \phi_{\mathrm{osc}}^{2n-4}
    \ll
    H_{\mathrm{osc}}^2
    \simeq
    m_\Phi^2,
    \label{eq:lambda}
\end{align}
so that the non-renormalizable term has a negligible effect on the dynamics of the AD field.
Since the field remains nearly constant until the onset of oscillation, we have
$\phi_{\mathrm{osc}}=\phi_i$ and $\theta_{\mathrm{osc}}=\theta_i$.
Their fluctuations are therefore given by Eq.~\eqref{eq:AD_fluctuation}.

During the oscillation, the amplitude of the AD field and its energy density evolve as
$\phi\propto a^{-3/2}$ and $\rho_{\mathrm{AD}}\propto a^{-3}$, respectively, where
$a$ is the scale factor.
Furthermore, when the AD field starts oscillating, it is kicked in the phase direction by the phase-dependent $A$-term and begins to rotate in the complex field plane. 
This rotation generates a baryon asymmetry because the angular momentum in the complex field plane corresponds to the baryon number.
The baryon number density is given by
\begin{equation}
    n_B=i\left(\dot{\Phi}^\ast \Phi - \Phi^\ast \dot{\Phi}\right),
\end{equation}
whose evolution is described by
\begin{align}
   \frac{1}{a^3}\frac{\partial}{\partial t}\left(n_B a^3\right)
   = \frac{1}{2i}\left(\Phi^*\frac{\partial V}{\partial\Phi^*}
       - \Phi\frac{\partial V}{\partial\Phi}\right)
       = \frac{na_M \lambda m_{3/2}}{M_\mathrm{pl}^{n-3}}
       |\Phi|^n\sin n\theta .
\end{align}
Therefore, the generated baryon number density is given by
\begin{align}
    a^3 n_B & = 
    \frac{na_M\lambda m_{3/2}}
    {2^{n/2}M_\mathrm{pl}^{n-3}}
    \int^t \mathrm{d}t'\,
    a^3(t')\phi^n\sin n\theta 
    \nonumber \\
    & \simeq
    \frac{na_M \lambda m_{3/2}}
    {2^{n/2}M_\mathrm{pl}^{n-3}}
    a_\mathrm{osc}^3
    \phi_\mathrm{osc}^n   
    H^{-1}_\mathrm{osc} \sin n\theta_\mathrm{osc} 
    \nonumber \\
    & \simeq
    a_\mathrm{osc}^3\frac{na_M \lambda m_{3/2}}
    {2^{n/2}m_\Phi M_\mathrm{pl}^{n-3}}
    \phi_\mathrm{osc}^n\sin n\theta_\mathrm{osc}~,
    \label{eq:baryon_number}
\end{align}
where we have used the fact that the baryon number is generated predominantly around
$H_{\mathrm{osc}}\simeq m_\Phi$.
This mechanism for generating the baryon asymmetry is known as the Affleck-Dine mechanism.

The baryon-to-entropy ratio, $\eta_B$, is then given by
\begin{equation}
    \label{eq:baryon_asymmetry}
    \eta_B = \frac{n_B}{s}
    = \frac{3T_R}{4}
    \left.\frac{n_B}{3H^2M_\mathrm{pl}^2}\right|_{\mathrm{osc}}
    \simeq
    \tilde{a}\frac{m_{3/2}T_R}{m_\Phi^3M_\mathrm{pl}^{n-1}}
    \phi_{\mathrm{osc}}^n\sin n\theta_\mathrm{osc} ,
\end{equation}
where $T_R$ is the reheating temperature.\footnote{
The curvaton decay may generate additional entropy and dilute $\eta_B$.
However, as shown below, the resulting entropy production is modest, and the dilution factor is smaller than two.
} 
We have assumed that reheating occurs after the onset of the AD-field oscillation. 
The coefficient $\tilde{a}$ is defined by
\begin{equation}
\tilde{a}
=
\frac{na_M \lambda}{2^{(n+4)/2}}.
\end{equation}

\subsection{Q-ball formtion}
\label{sec:Qball_formation}

It is well known that an oscillating AD field develops spatial instabilities when its potential is flatter than quadratic, leading to the formation of Q-balls.
Whether Q-balls form depends on the form of the SUSY-breaking potential, $V_{\mathrm{SB}}$.
In this paper, we consider a gauge-mediated SUSY-breaking scenario, in which the SUSY-breaking potential at large $|\Phi|$ is given by
\begin{equation}
    \label{eq:SB_potential}
    V_{\mathrm{SB}}(\Phi)
    =
    M_F^4
    \left[
    \log\left(\frac{|\Phi|^2}{M_{\mathrm{mess}}^2}\right)
    \right]^2
    +
    m_{3/2}^2|\Phi|^2
    \left[
    1
    +
    K\log\left(\frac{|\Phi|^2}{M_*^2}\right)
    \right],
\end{equation}
where $M_F$ denotes the SUSY-breaking scale and $M_{\mathrm{mess}}$ is the messenger mass.
The first and second terms in Eq.~\eqref{eq:SB_potential} represent the gauge-mediated and gravity-mediated contributions, respectively.
The latter is always present owing to gravitational interactions and includes a one-loop correction characterized by the renormalization scale $M_*$ and the dimensionless coefficient $K$.
The mass parameter $m_\Phi$ introduced in the previous section should be understood as the effective radial mass of the AD field at the onset of oscillation.\footnote{
More precisely,
$m_\Phi^2\equiv \left.\partial^2 V_{\mathrm{SB}}/\partial\phi^2 \right|_{\phi=\phi_{\mathrm{osc}}}$.
}

Which contribution dominates at the onset of oscillation determines the type of Q-ball that is formed. 
When $\phi_{\mathrm{osc}}>\phi_{\mathrm{eq}}\simeq M_F^2/m_{3/2}$, the gravity-mediated contribution dominates, and Q-balls are formed if $K<0$. 
On the other hand, when $\phi_{\mathrm{osc}}<\phi_{\mathrm{eq}}$, the gauge-mediated contribution dominates, and Q-balls are always formed. 
Q-balls formed in the gravity-mediated (gauge-mediated) regime are referred to as gravity-mediated-type (gauge-mediated-type) Q-balls.
If $K>0$ and $\phi_{\mathrm{osc}}>\phi_{\mathrm{eq}}$, Q-balls are formed only after the oscillation amplitude decreases below $\phi_{\mathrm{eq}}$. 
In this case, Q-ball formation is delayed relative to the onset of oscillation, and the resulting Q-balls are called delayed-type Q-balls.

In this paper, we consider the case with $\phi_{\mathrm{osc}}>\phi_{\mathrm{eq}}$. 
Depending on the sign of $K$, the resulting Q-balls are either gravity-mediated-type ($K<0$) or delayed-type ($K>0$).
The typical baryon charge and other properties of gravity-mediated-type Q-balls are given by
\begin{align}
    Q & \simeq 1.9\times10^{-2}
    \left(\frac{\phi_\mathrm{osc}}{m_{3/2}}\right)^2, \\
    M_Q & \simeq m_{3/2}Q, \\
    R_Q & \simeq |K|^{-1/2}m_{3/2}^{-1}, \\
    \omega_Q & \simeq m_{3/2},
\end{align}
where $M_Q$ and $R_Q$ are the mass and radius of the Q-ball, respectively, and $\omega_Q=dM_Q/dQ$.
On the other hand, for delayed-type Q-balls, the corresponding quantities are given by
\begin{align}
    Q & \simeq 6\times10^{-4}
    \left(\frac{M_F}{m_{3/2}}\right)^4, \\
    M_Q & \simeq \frac{4\sqrt{2}\pi}{3}\zeta M_F Q^{3/4}, \\
    R_Q & \simeq \frac{1}{\sqrt{2}\zeta}M_F^{-1}Q^{1/4}, \\
    \omega_Q & \simeq \sqrt{2}\pi\zeta M_F Q^{-1/4},
\end{align}
where $\zeta=\mathcal{O}(1)$.

\subsection{Q-ball decay}
\label{sec:Qball_decay}

Since Q-balls behave as non-relativistic matter after their formation, their energy density evolves as $\rho_Q\propto a^{-3}$.
Therefore, the Q-ball energy density fraction increases during the radiation-dominated era after reheating, and eventually the Q-balls decay into radiation.
In the AD curvaton scenario, the AD-field condensate (i.e., Q-balls) constitutes a significant fraction of the cosmic energy density at the time of its decay.

The decay rate of a Q-ball is given by
\begin{equation}
    \Gamma_Q \simeq
    \frac{N_q}{Q}
    \frac{\omega_Q^3}{12\pi^2}
    4\pi R_Q^2,
    \label{eq:decay}
\end{equation}
where $N_q$ denotes the effective number of decay channels, and we take $N_q=18$.
The decay temperature $T_{\mathrm{dec}}$ is estimated from the condition
$\Gamma_Q\simeq H(T_{\mathrm{dec}})$ as
\begin{equation}
    T_{\mathrm{dec}}
    \simeq
    \left(
    \frac{90}{\pi^2 g_{\mathrm{dec}}}
    \right)^{1/4}
    \sqrt{\Gamma_Q M_{\mathrm{pl}}},
    \label{eq:decay_temp}
\end{equation}
where $g_{\mathrm{dec}}$ is the effective number of relativistic degrees of freedom at $T_{\mathrm{dec}}$.
Hereafter, the subscript “dec” denotes
quantities evaluated at the decay of the Q-balls.

Here, we evaluate the relation between the decay temperature, $T_\mathrm{dec}$, and the reheating temperature, $T_R$.
The energy density of the AD field evolves from the onset of oscillation to the time of decay according to
\begin{align}
    \rho_{\mathrm{AD},\mathrm{osc}}
    &=
    \left(\frac{a_R}{a_\mathrm{osc}}\right)^3
    \left(\frac{a_\mathrm{dec}}{a_R}\right)^3
    \rho_{\mathrm{AD},\mathrm{dec}}
    \nonumber\\
    &=
    \frac{m_{3/2}^2}
    {
        \dfrac{1}{3M_\mathrm{pl}^2}
        \dfrac{\pi^2}{30}g_R T_R^4
    }\,
    \frac{g_R T_R^3}
         {g_\mathrm{dec}T_{I,\mathrm{dec}}^3}
    \rho_{\mathrm{AD},\mathrm{dec}},
\end{align}
where $T_{I,\mathrm{dec}}$ is the temperature of the radiation produced by inflaton decay immediately before the Q-ball decay.
Defining $r_\mathrm{dec}$ by
\begin{equation}
    r_\mathrm{dec}
    =
    \frac{\rho_{\mathrm{AD},\mathrm{dec}}}
         {\rho_{I,\mathrm{dec}}},
\end{equation}
we can express $T_{I,\mathrm{dec}}$ as
\begin{equation}
    T_{I,\mathrm{dec}}
    =
    \frac{T_\mathrm{dec}}
         {(1+r_\mathrm{dec})^{1/4}},
\end{equation}
where $\rho_{I,\mathrm{dec}}$ is the energy density of the radiation produced by inflaton decay immediately before the Q-ball decay.

Since the AD-field energy density at the onset of oscillation is given by
\begin{align}
    \rho_{\mathrm{AD},\mathrm{osc}}
    \simeq
    \frac{1}{2}m_{3/2}^2\phi_\mathrm{osc}^2,
\end{align}
we obtain
\begin{equation}
    T_R
    \simeq
    \frac{6r_\mathrm{dec}}
         {(1+r_\mathrm{dec})^{1/4}}\,
    T_\mathrm{dec}
    \left(
        \frac{\phi_\mathrm{osc}}{M_\mathrm{pl}}
    \right)^{-2}.
    \label{eq:reheating_temp}
\end{equation}

\subsection{Curvature perturbations}
\label{sec:curvature_perterbation}

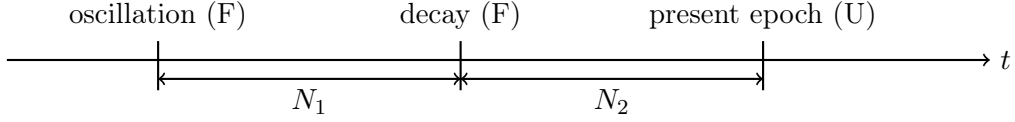
\begin{figure}[t]
    \centering
    \begin{tikzpicture}[node distance=1.5cm, auto]
      \draw[->,thick] (-6,0) -- (7,0) node[right] {$t$};
      \draw[thick] (-4,-0.4) -- (-4,0.25) node[above] {oscillation (F)};
      \draw[thick] (0,-0.4) -- (0,0.25) node[above] {decay (F)};
      \draw[thick] (4,-0.4) -- (4,0.25) node[above] {present epoch (U)};
      \draw[<->,thick] (-4,-0.25) -- (0,-0.25) node[midway, below] {$N_1$};
      \draw[<->,thick] (0,-0.25) -- (4,-0.25) node[midway, below] {$N_2$};
    \end{tikzpicture}
    \caption{Schematic timeline used in the $\delta N$ calculation for the AD curvaton scenario.
    The onset of AD-field oscillation and the Q-ball decay are evaluated on spatially flat hypersurfaces, denoted by ``F,'' while the final hypersurface corresponding to the present epoch is a uniform-density hypersurface, denoted by ``U.''
    The quantities $N_1$ and $N_2$ represent the numbers of e-folds between these hypersurfaces.}
    \label{fig:timeline}
\end{figure}

Since the AD field acquires fluctuations during inflation, the decay of the AD-field condensate generates curvature perturbations. 
We calculate the curvature perturbations generated by the AD curvaton following Ref.~\cite{Kohri:2013qwa}.
To calculate the curvature perturbations, we employ the $\delta N$ formalism, choosing the uniform-density slice (U) as the final hypersurface for the present observer and the flat slice (F) as the initial hypersurface at both the onset of oscillation and the time of decay, as illustrated in Fig.~\ref{fig:timeline}.

Since we assume that the inflaton generates negligible curvature perturbations, we only need to consider the number of $e$-folds after the onset of oscillation.
In the present scenario, $\delta N$ can be expanded as
\begin{equation}
    \delta N
    =
    \frac{dN}{d\phi_\mathrm{osc}}
    \delta\phi_\mathrm{osc}
    +
    \frac12
    \frac{d^2N}{d\phi_\mathrm{osc}^2}
    (\delta\phi_\mathrm{osc})^2
    +\cdots .
\end{equation}
The first term gives the linear curvature perturbation, while the second term gives rise to primordial non-Gaussianity.
We first consider the number of $e$-folds, $N_1$, between the onset of oscillation and the decay of the AD-field condensate.
We define
\begin{align}
  X \equiv e^{N_1}
  =
  \frac{a_\mathrm{dec}}{a_\mathrm{osc}}.
\end{align}
For the radiation component produced by the inflaton, we have
\begin{align}
   X^4
   =
   \frac{\rho_{I,\mathrm{osc}}}{\rho_{I,\mathrm{dec}}}
   =
   \frac{3M_\mathrm{pl}^2m_\Phi^2}
   {3M_\mathrm{pl}^2\Gamma^2
   -\frac{1}{2}m_\Phi^2\phi_\mathrm{osc}^2X^{-3}},
\end{align}
which yields the quartic equation
\begin{align}
  3 M_\mathrm{pl}^2\Gamma^2X^4
  -\frac{1}{2}m_\Phi^2\phi_\mathrm{osc}^2X
  =
  3M_\mathrm{pl}^2m_\Phi^2 .
  \label{eq:relation_X}
\end{align}
Differentiating both sides of Eq.~\eqref{eq:relation_X} with respect to $\phi_\mathrm{osc}$, we obtain
\begin{align}
   6 M_\mathrm{pl}^2\Gamma\frac{d\Gamma}{d\phi_\mathrm{osc}} X^4
   +12 M_\mathrm{pl}^2\Gamma^2X^3\frac{dX}{d\phi_\mathrm{osc}}
   -m_\Phi^2\phi_\mathrm{osc}X
   -\frac{1}{2}m_\Phi^2\phi_\mathrm{osc}^2\frac{dX}{d\phi_\mathrm{osc}}
   =0 ,
\end{align}
or equivalently,
\begin{align}
   \frac{1}{X}\frac{dX}{d\phi_\mathrm{osc}}
   =
   \frac{
      m_\Phi^2\phi_\mathrm{osc}X
      -6M_\mathrm{pl}^2\Gamma
      \dfrac{d\Gamma}{d\phi_\mathrm{osc}}X^4
   }{
      12M_\mathrm{pl}^2\Gamma^2X^4
      -\dfrac12m_\Phi^2\phi_\mathrm{osc}^2X
   }~.
   \label{eq:diff_relation_X}
\end{align}
Defining 
$f_\mathrm{dec}$ by
\begin{align}
    f_\mathrm{dec}
    &\equiv
    \frac{3r_\mathrm{dec}}
         {4+3r_\mathrm{dec}}
    =
    \frac{
      \frac32
      m_\Phi^2
      \phi_\mathrm{osc}^2
      X^{-3}
    }{
      12M_\mathrm{pl}^2\Gamma^2
      -
      \frac12
      m_\Phi^2
      \phi_\mathrm{osc}^2
      X^{-3}
    },
    \label{eq:fdec}
\end{align}
Eq.~\eqref{eq:diff_relation_X} can then be rewritten as
\begin{align}
    \frac{dN_1}{d\phi_\mathrm{osc}}
    =
    \frac{2f_\mathrm{dec}}
         {3\phi_\mathrm{osc}}
    -
    \left(
       \frac12
       +
       \frac{f_\mathrm{dec}}6
    \right)
    \frac1\Gamma
    \frac{d\Gamma}{d\phi_\mathrm{osc}}.
\end{align}

Next, we consider the number of $e$-folds, $N_2$, from the decay epoch to the present time.
It is given by
\begin{equation}
    N_2
    =
    \ln\frac{a_0}{a_\mathrm{dec}}
    =
    \ln\frac{a_\mathrm{eq}}{a_\mathrm{dec}}
    +
    \ln\frac{a_0}{a_\mathrm{eq}},
\end{equation}
where $a_\mathrm{eq}$ is the scale factor at matter-radiation equality.
Since $H\propto a^{-2}$ during the radiation-dominated era, we obtain
\begin{equation}
    \frac{dN_2}{d\phi_\mathrm{osc}}
    =
    -\frac{d\ln a_\mathrm{dec}}{d\phi_\mathrm{osc}}
    =
    \frac{1}{2}\frac{d\ln H_\mathrm{dec}}{d\phi_\mathrm{osc}}
    =
    \frac{1}{2\Gamma}\frac{d\Gamma}{d\phi_\mathrm{osc}}.
\end{equation}
where we have used $H_\mathrm{dec}=\Gamma$.
For the total number of $e$-folds, $N=N_1+N_2$, we obtain
\begin{align}
    \frac{dN}{d\phi_\mathrm{osc}}
    =
    \frac{dN_1}{d\phi_\mathrm{osc}}+\frac{dN_2}{d\phi_\mathrm{osc}}
    =
    \frac{2f_\mathrm{dec}}{3\phi_\mathrm{osc}}
    -
    \frac{f_\mathrm{dec}}{6}\frac{1}{\Gamma}\frac{d\Gamma}{d\phi_\mathrm{osc}},
    \label{eq:dN_dphi}
\end{align}
from which the curvature perturbation is obtained as
\begin{align}
    \zeta
    =
    \delta N
    =
    \frac{2}{3}
    f_\mathrm{dec}
    \frac{\delta\phi_\mathrm{osc}}
         {\phi_\mathrm{osc}}
    -
    \frac{1}{6}
    f_\mathrm{dec}
    \frac{1}{\Gamma}
    \frac{d\Gamma}{d\phi_\mathrm{osc}}
    \delta\phi_\mathrm{osc}.
\end{align}
In the present scenario, the decay rate is determined by Q-ball decay and scales as
$\Gamma\propto\phi_{\mathrm{osc}}^{\alpha}$,
where $\alpha=-2$ for gravity-mediated-type Q-balls and $\alpha\simeq0$ for delayed-type Q-balls.\footnote{%
This scaling remains unchanged even when the charge distribution of the formed Q-balls is taken into account, because the distribution can be expressed as a function of $Q/Q_*$, where $Q_*$ denotes the typical charge of the formed Q-balls~\cite{Hiramatsu:2010dx,Kasuya:2025nix}.
}
Therefore, we obtain
\begin{equation}
    \label{eq:curvature_perturbation}
    \zeta = \left(\frac{2}{3}-\frac{\alpha}{6}\right)f_\mathrm{dec}\frac{\delta\phi_\mathrm{osc}}{\phi_\mathrm{osc}}~.
\end{equation}
The CMB observations indicate that the amplitude of the curvature perturbation power spectrum is
$\mathcal{P}_\zeta(k_*)\simeq2.1\times10^{-9}$,
where $k_*=0.05\,\mathrm{Mpc}^{-1}$ is the pivot scale.
Since $\delta\phi_\mathrm{osc}\simeq\delta\phi_i=H_I/(2\pi)$ and
$\phi_\mathrm{osc}\simeq\phi_i$, the CMB normalization requires
\begin{equation}
    \label{eq:phi_osc}
    \phi_\mathrm{osc}
    \simeq
    3.47\times10^3
    \left(
        \frac{2}{3}-\frac{\alpha}{6}
    \right)
    f_\mathrm{dec}
    H_I.
\end{equation}
Substituting Eq.~\eqref{eq:phi_osc} into Eq.~\eqref{eq:baryon_asymmetry}, we obtain
\begin{equation}
\label{eq:baryon_asymmetry_our_model}
   \begin{aligned}
    \eta_B 
    & \simeq
    10^{-10}
    \left(\frac{m_{3/2}}{10\,\mathrm{GeV}}\right)^{-2}
        \left(\frac{T_R}{10^6\,\mathrm{GeV}}\right)
    \\[0.7em]
    & \times 
    \begin{cases}
        \left(\frac{\tilde{a}}{0.6\times 10^{-19}}\right)
        \left(\frac{2}{3}-\frac{\alpha}{6}\right)^4
        \left(\frac{f_\mathrm{dec}}{0.2}\right)^4
        \left(\frac{H_I}{10^{12}\,\mathrm{GeV}}\right)^4
        \sin 4\theta_\mathrm{osc}
        & (n=4)
        \\[0.7em]
        \left(\frac{\tilde{a}}{0.8\times10^{-11}}\right)
        \left(\frac{2}{3}-\frac{\alpha}{6}\right)^6
        \left(\frac{f_\mathrm{dec}}{0.2}\right)^6
        \left(\frac{H_I}{10^{12}\,\mathrm{GeV}}\right)^6
        \sin 6\theta_\mathrm{osc}
        & (n=6)~.
    \end{cases}
    \end{aligned}
\end{equation}
Here we have assumed $m_{3/2}=m_\Phi$. 
Therefore, the observed baryon asymmetry and curvature perturbations can be simultaneously reproduced for a suitable choice of $H_I$ and $\tilde{a}$.

Let us estimate the non-Gaussianity of the curvature perturbations generated in the curvaton scenario.
The non-Gaussianity is characterized by the nonlinearity parameter $f_{\mathrm{NL}}$ through
\begin{equation}
    \zeta
    =
    \zeta_G
    +
    \frac{3}{5}
    f_{\mathrm{NL}}
    \left(
        \zeta_G^2
        -
        \langle\zeta_G^2\rangle
    \right),
\end{equation}
where $\zeta_G$ denotes the Gaussian part of the curvature perturbation. The nonlinearity parameter is given by
\begin{equation}
    f_{\mathrm{NL}}
    =
    \frac{5}{6}
    \left(
        \frac{dN}{d\phi_\mathrm{osc}}
    \right)^{-2}
    \frac{d^2N}{d\phi_\mathrm{osc}^2}.
\end{equation}
Differentiating Eq.~\eqref{eq:dN_dphi} with respect to $\phi_\mathrm{osc}$, we obtain
\begin{align}
    \frac{d^2N}{d\phi_\mathrm{osc}^2}
    = &
    \frac{dN}{d\phi_\mathrm{osc}}
    \frac{1}{f_\mathrm{dec}}
    \frac{df_\mathrm{dec}}{d\phi_\mathrm{osc}}
    -
    \frac{2f_\mathrm{dec}}{3\phi_\mathrm{osc}^2}
    +
    \frac{f_\mathrm{dec}}{6}
    \frac{1}{\Gamma^2}
    \left(
        \frac{d\Gamma}{d\phi_\mathrm{osc}}
    \right)^2
    -
    \frac{f_\mathrm{dec}}{6}
    \frac{1}{\Gamma}
    \frac{d^2\Gamma}{d\phi_\mathrm{osc}^2}
    \nonumber\\
    = &
    \frac{3-2f_\mathrm{dec}-f_\mathrm{dec}^2}
         {f_\mathrm{dec}}
    \left(
        \frac{dN}{d\phi_\mathrm{osc}}
    \right)^2
    -
    \frac{2f_\mathrm{dec}}{3\phi_\mathrm{osc}^2}
    +
    \frac{f_\mathrm{dec}}{6}
    \frac{1}{\Gamma^2}
    \left(
        \frac{d\Gamma}{d\phi_\mathrm{osc}}
    \right)^2
    -
    \frac{f_\mathrm{dec}}{6}
    \frac{1}{\Gamma}
    \frac{d^2\Gamma}{d\phi_\mathrm{osc}^2}.
\end{align}
Substituting this relation into the above expression, we obtain
\begin{align}
    f_{\mathrm{NL}}
    =
    \frac{5}{6}
    \left(
        \frac{dN}{d\phi_\mathrm{osc}}
    \right)^{-2}
    \frac{d^2N}{d\phi_\mathrm{osc}^2}
    =
    \frac{5}{6}
    \left(
        -2
        -
        f_\mathrm{dec}
        +
        \frac{3(2-\alpha)}
             {f_\mathrm{dec}(4-\alpha)}
    \right)~,
    \label{eq:nonlinearity}
\end{align}
where we have used $\Gamma\propto\phi_\mathrm{osc}^{\alpha}$

\section{Isocurvature perturbations}
\label{sec:iso_curv_perterbation}

We have shown that the AD field can play the role of a curvaton and generate curvature perturbations. 
In the AD curvaton scenario, the AD field also generates the baryon asymmetry of the Universe [see Eq.~\eqref{eq:baryon_asymmetry_our_model}]. 
However, this leads to a cosmological problem because fluctuations of the AD field generate baryon isocurvature perturbations, which are stringently constrained by CMB observations. 
In this section, we discuss the baryon isocurvature perturbation problem and present a possible solution.

From Eq.~\eqref{eq:baryon_number}, the fluctuation in the generated baryon number density is given by
\begin{equation}
    \frac{\delta n_B}{n_B}
    =
    n\frac{\delta\phi_\mathrm{osc}}{\phi_\mathrm{osc}}
    +
    n\cot n\theta_\mathrm{osc},\delta\theta_\mathrm{osc}.
\end{equation}
Since the first term is proportional to the curvature perturbation [Eq.~\eqref{eq:curvature_perturbation}], it gives rise to correlated baryon isocurvature perturbations. 
On the other hand, the second term is independent of the curvature perturbation and gives rise to uncorrelated baryon isocurvature perturbations.
In the AD curvaton scenario, the correlated baryon isocurvature perturbations are unavoidable because $\delta\phi_\mathrm{osc}\neq0$ is required to generate the observed curvature perturbations. 
Therefore, we first discuss the correlated baryon isocurvature perturbations.

\subsection{Correlated isocurvature perturbations}
\label{sec:correlated_iso_pert}

The correlated baryon isocurvature perturbation, $S_{B,\,\mathrm{corr}}$, is given by
\begin{equation}
    \label{eq:corr_baryon_iso}
    S_{B,\,\mathrm{corr}}
    =
    n\frac{\delta\phi_\mathrm{osc}}{\phi_\mathrm{osc}}-3\zeta
    =
    \left(n-2f_\mathrm{dec}+\frac{\alpha}{2}f_\mathrm{dec}\right)
    \frac{\delta\phi_\mathrm{osc}}{\phi_\mathrm{osc}}.
\end{equation}
This induces the correlated matter isocurvature perturbation,
\begin{equation}
    \label{eq:corr_matter_iso}
    S_{m,\,\mathrm{corr}}
    =\frac{\Omega_B}{\Omega_m}\,S_{B,\,\mathrm{corr}}
    \simeq 0.16 S_{B,\,\mathrm{corr}},
\end{equation}
where $\Omega_B$ and $\Omega_m$ are the baryon and matter density parameters, respectively.
From Eqs.~\eqref{eq:curvature_perturbation} and \eqref{eq:corr_matter_iso}, we find that $S_{m,\,\mathrm{corr}}=\mathcal{O}(\zeta)$.
This is inconsistent with the CMB constraint~\cite{Planck:2018jri},
\begin{equation}
    \frac{\mathcal{P}_{S_{m,\,\mathrm{corr}}}(k_*)}
    {\mathcal{P}_\zeta(k_*)}<0.00095,
    \label{eq:corr_iso}
\end{equation}
where $\mathcal{P}_i$ denotes the power spectrum of the perturbation $i$.
Therefore, the correlated matter isocurvature perturbation must be suppressed.
Here, it is worth noting that $S_{B,\,\mathrm{corr}}$ would be suppressed if $n=2$, $f_\mathrm{dec}\simeq1$, and Q-ball formation did not occur. 
However, in the SUSY Affleck-Dine mechanism, one always has $n\ge4$, and therefore $S_{B,\,\mathrm{corr}}$ cannot be suppressed in this way.

\begin{figure}[t]
  \centering
  \includegraphics[width=0.99\linewidth]{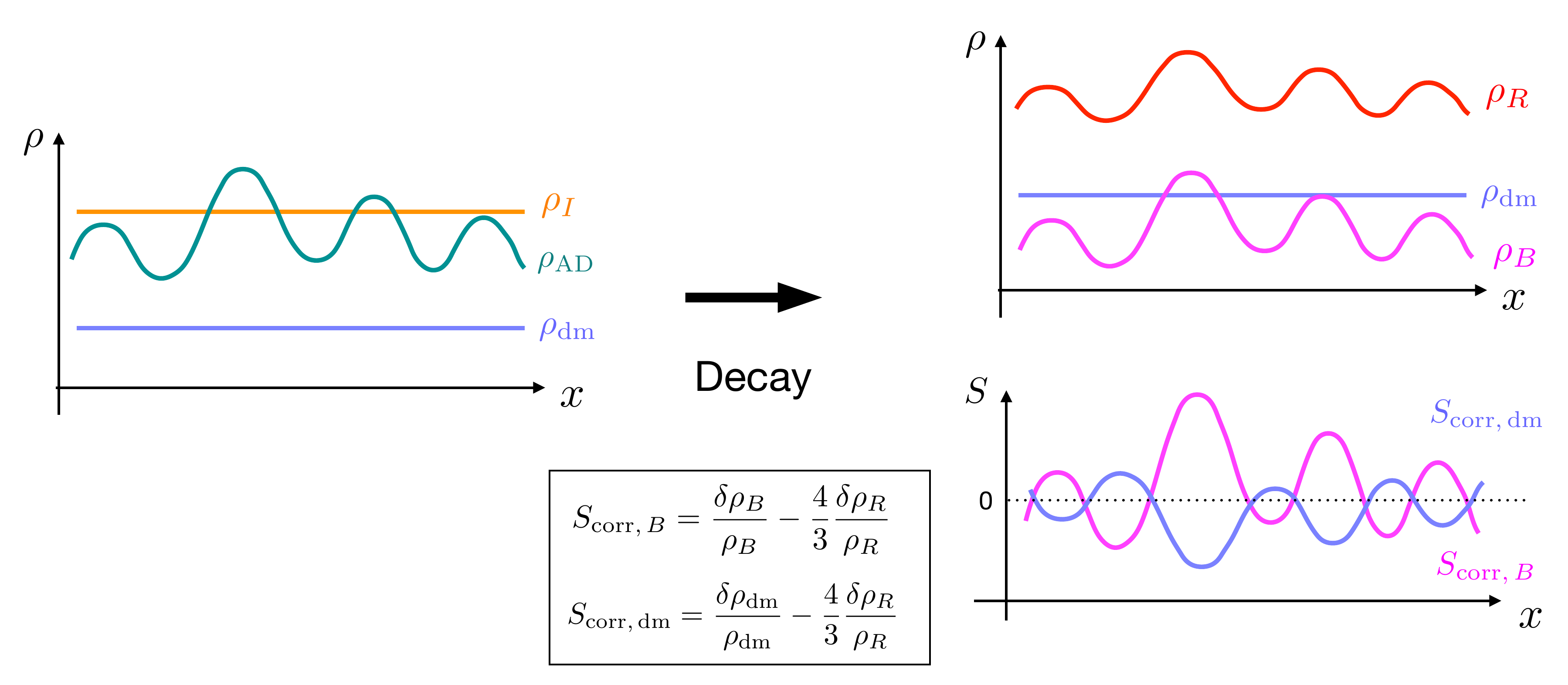}
  \caption{
  Schematic illustration of the energy-density fluctuations and correlated isocurvature perturbations.
  The left panel shows the fluctuations in the energy densities of the AD field, $\rho_{\mathrm{AD}}$, the radiation originating from inflaton decay, $\rho_I$, and dark matter, $\rho_{\mathrm{dm}}$, before the curvaton decay.
  The upper-right panel shows the fluctuations in the radiation, baryon, and dark matter energy densities, $\rho_R$, $\rho_B$, and $\rho_{\mathrm{dm}}$, respectively, after the curvaton decay.
  The lower-right panel shows the resulting correlated baryon and dark matter isocurvature perturbations.
  }
  \label{fig:corr_isocurv}
\end{figure}

The correlated matter isocurvature perturbation can be suppressed if there exists an anti-correlated dark matter isocurvature perturbation.
Suppose that dark matter is produced before the curvaton decays and decouples from the thermal bath thereafter. 
Since the inflaton generates negligible curvature perturbations, the dark matter is initially unperturbed. 
After the AD field decays and generates the curvature perturbations, the dark matter acquires an anti-correlated isocurvature perturbation,
\begin{equation}
    S_{\mathrm{dm,\,corr}}
    =
    \frac{\delta\rho_\mathrm{dm}}{\rho_\mathrm{dm}}
    -3\zeta
    =
    -3\zeta,
\end{equation}
where $\rho_\mathrm{dm}$ and $\delta\rho_\mathrm{dm}$ are the dark matter energy density and its perturbation, respectively.
Including the contribution from dark matter, the total correlated matter isocurvature perturbation is given by
\begin{align}
    S_{m,\,\mathrm{corr}}
    &=
    \frac{\Omega_B}{\Omega_m}\,
    S_{B,\,\mathrm{corr}}
    +
    \frac{\Omega_{\mathrm{dm}}}{\Omega_m}\,
    S_{\mathrm{dm,\,corr}}
    \nonumber\\
    &=
    \left[
        \left(
            n
            -2f_\mathrm{dec}
            +\frac{\alpha}{2}f_\mathrm{dec}
        \right)
        \frac{\Omega_B}{\Omega_m}
        +\left(-2f_\mathrm{dec}
            +\frac{\alpha}{2}f_\mathrm{dec}\right)
        \frac{\Omega_\mathrm{dm}}{\Omega_m}
    \right]
    \frac{\delta\phi_\mathrm{osc}}
         {\phi_\mathrm{osc}}.
    \label{eq:total_matter_corr_iso}
\end{align}
Therefore, the stringent CMB constraint can be evaded if the quantity in square brackets in Eq.~\eqref{eq:total_matter_corr_iso} is approximately zero. 
This condition leads to
\begin{equation}
    f_\mathrm{dec}
    \simeq\frac{2 n}{4-\alpha}\frac{\Omega_B}{\Omega_m}.
    \label{eq:compensation_fdec}
\end{equation}
Thus, the present scenario realizes compensated isocurvature perturbations~\cite{Holder:2009gd,Gordon:2009wx,Grin:2011tf,Grin:2013uya,Harigaya:2014bsa}.
Figure~\ref{fig:corr_isocurv} schematically illustrates the fluctuations in the energy densities of the various components and the resulting correlated isocurvature perturbations.

Now let us discuss the dark matter candidate in the present scenario.
In gauge-mediated SUSY-breaking models, the gravitino is most likely the lightest supersymmetric particle (LSP) and is therefore a natural dark matter candidate. 
Gravitinos are thermally produced during reheating after inflation and subsequently decouple from the thermal bath. 
Therefore, gravitino dark matter is an ideal candidate for generating anti-correlated isocurvature perturbations.
The gravitino density is estimated as~\cite{Kawasaki:2017bqm}
\begin{align}
    \Omega_{3/2}h^2
    \simeq
    0.35\left(\frac{m_{3/2}}{10\,\mathrm{GeV}}\right)^{-1}
    \left(\frac{m_{\tilde{g}}}{10^4\,\mathrm{GeV}}\right)^2
    \left(\frac{T_R}{10^6\,\mathrm{GeV}}\right),
    \label{eq:gravitino}
\end{align}
where $h$ is the present Hubble parameter in units of 100\,km/s/Mpc, $T_R$ is the reheating temperature and $m_{\tilde{g}}$ is the gluino mass.

\subsection{Uncorrelated isocurvature perturbations}
\label{sec:uncorrelated_iso_pert}

Let us discuss the uncorrelated baryon isocurvature perturbation $S_{B,\,\mathrm{uncorr}}$ in this scenario, which is given by
\begin{equation}
    S_{B,\,\mathrm{uncorr}}= n\cot n\theta_\mathrm{osc},\delta\theta_\mathrm{osc}
    \simeq n\cot n\theta_\mathrm{osc}\,\frac{H_I}{2\pi \phi_\mathrm{osc}},
\end{equation}
where we have used Eq.~\eqref{eq:AD_fluctuation} with $\phi_i \simeq \phi_\mathrm{osc}$.
This induces the uncorrelated matter isocurvature perturbation 
\begin{equation}
    \label{eq:uncorr_matter_iso}
    S_{m,\,\mathrm{uncorr}} \simeq \frac{\Omega_B}{\Omega_m}\, 
    \left(n\cot n\theta_\mathrm{osc}\right)\,
    \frac{H_I}{2\pi \phi_\mathrm{osc}}~.
\end{equation}
Using Eqs.~\eqref{eq:curvature_perturbation} and \eqref{eq:uncorr_matter_iso} we obtain
\begin{equation}
    \frac{\mathcal{P}_{S_{m,\,\mathrm{uncorr}}}(k_*)}{\mathcal{P}_\zeta(k_*)}
    = \left(\frac{\Omega_B}{\Omega_m}\right)^2
       \frac{(n\cot n\theta_\mathrm{osc})^2}{f_\mathrm{dec}^2}~,
\end{equation}
which should satisfies the observational constraint~\cite{Planck:2018jri} given by
\begin{equation}
    \frac{\mathcal{P}_{S_{m,\,\mathrm{uncorr}}}(k_*)}
    {\mathcal{P}_\zeta(k_*)}<0.0395~.
    \label{eq:uncor_obs_constr}
\end{equation}
To satisfy this constraint we should require $n\theta_\mathrm{osc} \sim \pi/2$.

\section{Conditions for a successful Affleck-Dine curvaton scenario}
\label{sec:successful_scenario}

In this section, we derive the conditions under which the AD curvaton scenario successfully generates both curvature perturbations and the baryon asymmetry of the Universe without inducing excessively large isocurvature perturbations.
Since the AD field can form two types of Q-balls---gravity-mediated and delayed-type Q-balls---we discuss each case separately.
Hearafter, we take $n=6$ for simplicity.

\subsection{The gravity-mediated-type Q-ball scenario}

In the gravity-mediated-type Q-ball scenario, the decay rate depends on the AD-field value with $\alpha=-2$.
From Eq.~\eqref{eq:compensation_fdec}, the cancellation of the correlated isocurvature perturbations requires
\begin{align}
    f_\mathrm{dec}
    \simeq
    0.314.
\end{align}
The CMB constraint on the correlated isocurvature perturbations, given in Eq.~\eqref{eq:corr_iso}, requires
\begin{align}
    \left|\delta S_m\right|
    =
    \left|
        -3\frac{\delta f_\mathrm{dec}}{f_\mathrm{dec}}
    \right|
    \lesssim
    0.031,
    \label{eq:corr_iso_const}
\end{align}
which corresponds to a fine-tuning of $f_\mathrm{dec}$ at the $1\%$ level.
The CMB constraint on the uncorrelated isocurvature perturbations, given in Eq.~\eqref{eq:uncor_obs_constr}, requires an additional fine-tuning of approximately $4\%$:
\begin{align}
    1.505
    <
    6\theta_\mathrm{osc}
    <
    1.637.
\end{align}
Equation~\eqref{eq:phi_osc} determines the AD-field value as
\begin{align}
    \phi_\mathrm{osc}
    \simeq
    6.5\times10^{14}\,\mathrm{GeV}
    \left(
        \frac{H_I}{6\times10^{11}\,\mathrm{GeV}}
    \right).
    \label{eq:phi_osc_Hubble_relation}
\end{align}
The nonlinearity parameter estimated from Eq.~\eqref{eq:nonlinearity} is
\begin{align}
    f_\mathrm{NL}
    \simeq
    3.38.
\end{align}
This value is consistent with the CMB constraint
$f_\mathrm{NL}=-0.9\pm5.1$ at the $1\sigma$ level~\cite{Planck:2019kim}.

We can now determine the model parameters for which the AD curvaton scenario is successfully realized.
Using Eq.~\eqref{eq:phi_osc_Hubble_relation} together with Eqs.~\eqref{eq:decay_temp}, \eqref{eq:reheating_temp}, and \eqref{eq:gravitino}, we obtain $T_\mathrm{dec}$, $T_R$, and $m_{3/2}$ as
\begin{align}
    T_\mathrm{dec}
    &\simeq
    36\,\mathrm{MeV}
    \left(\frac{|K|}{0.01}\right)
    \left(\frac{g_\mathrm{dec}}{200}\right)^{1/2}
    \left(\frac{M_{\tilde{g}}}{10^4\,\mathrm{GeV}}\right)^{-6}
    \left(\frac{H_I}{6\times10^{11}\,\mathrm{GeV}}\right)^8,
    \\[0.6em]
    T_R
    &\simeq
    1.6\times10^6\,\mathrm{GeV}
    \left(\frac{|K|}{0.01}\right)
    \left(\frac{g_\mathrm{dec}}{200}\right)^{1/2}
    \left(\frac{M_{\tilde{g}}}{10^4\,\mathrm{GeV}}\right)^{-6}
    \left(\frac{H_I}{6\times10^{11}\,\mathrm{GeV}}\right)^6,
    \\[0.6em]
    m_{3/2}
    &\simeq
    47\,\mathrm{GeV}
    \left(\frac{|K|}{0.01}\right)
    \left(\frac{g_\mathrm{dec}}{200}\right)^{1/2}
    \left(\frac{M_{\tilde{g}}}{10^4\,\mathrm{GeV}}\right)^{-4}
    \left(\frac{H_I}{6\times10^{11}\,\mathrm{GeV}}\right)^6.
\end{align}

\begin{figure}[t]
  \centering
  \includegraphics[width=0.9\linewidth]{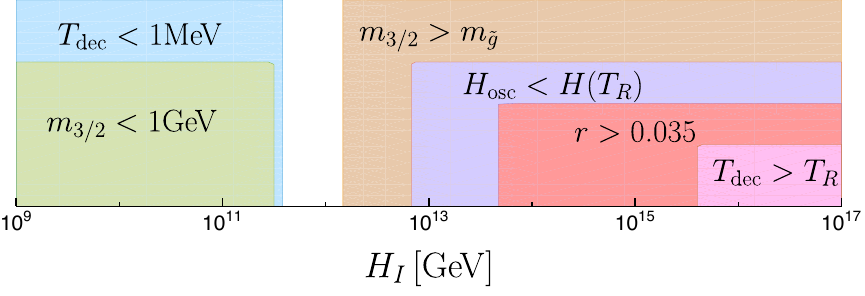}
  \caption{
  Constraints on the Hubble parameter $H_I$ during inflation.
  The green range corresponds to $m_{3/2}<1\,\mathrm{GeV}$, for which Q-balls cannot decay into protons.
  The blue range is excluded by the BBN constraint, $T_{\mathrm{dec}}<1\,\mathrm{MeV}$.
  The red range is excluded by the observational bound on the tensor-to-scalar ratio, $r>0.035$.
  The brown range corresponds to $m_{3/2}>M_{\tilde g}$ and is excluded because the gravitino, assumed to be the LSP, would be heavier than the gluino.
  The purple range corresponds to $H_{\mathrm{osc}}<H(T_R)$, implying that the AD field begins to oscillate after reheating.
  The magenta range corresponds to $T_{\mathrm{dec}}>T_R$, for which Q-ball decay occurs before the completion of reheating.
  The white range satisfies all the constraints.
  We take $M_{\tilde g}=10^4\,\mathrm{GeV}$.
  }
  \label{fig:const_hubble}
\end{figure}

There are several constraints on $H_I$, $T_{\mathrm{dec}}$, $T_R$, and $m_{3/2}$.
First, the decay temperature must satisfy
$T_{\mathrm{dec}}\gtrsim1\,\mathrm{MeV}$;
otherwise, Q-ball decay would spoil the successful predictions of BBN~\cite{Kawasaki:1999na,Kawasaki:2000en,Hannestad:2004px,Hasegawa:2019jsa}.
Second, Q-ball decay must occur after reheating, which requires
$T_R>T_{\mathrm{dec}}$.
Third, the gravitino mass must satisfy
$m_{3/2}\gtrsim1\,\mathrm{GeV}$
for Q-balls to be kinematically able to decay into protons.
Fourth, since we assume that the AD field begins to oscillate before reheating, the condition
$H_{\mathrm{osc}}\simeq m_{3/2}\gtrsim H(T_R)$
places the following upper bound on the reheating temperature:
\begin{equation}
    T_R
    \lesssim
    2.3\times10^9\,\mathrm{GeV}
    \left(\frac{|K|}{0.01}\right)^{1/2}
    \left(\frac{M_{\tilde{g}}}{10^4\,\mathrm{GeV}}\right)^{-2}
    \left(\frac{H_I}{6\times10^{11}\,\mathrm{GeV}}\right)^3.
\end{equation}
Fifth, since the gravitino is assumed to be the LSP, it must be lighter than the gluino: $m_{3/2}\lesssim M_{\tilde{g}}$.
Finally, the CMB constraint on the tensor-to-scalar ratio,
$r<0.035$~\cite{BICEP:2021xfz},
imposes the upper bound
$H_I\lesssim4.7\times10^{13}\,\mathrm{GeV}$.
These constraints are shown in Fig.~\ref{fig:const_hubble}
. 
From the figure, it is found that constraints can be simultaneously satisfied in the range
\begin{equation}
    3\times10^{11}\,\mathrm{GeV}
    \lesssim
    H_I
    \lesssim
    10^{12}\,\mathrm{GeV}.
\end{equation}

The condition in Eq.~\eqref{eq:lambda}, under which the non-renormalizable
term has a negligible effect on the AD-field dynamics, is satisfied provided that
\begin{align}
    \lambda
    \ll
    3.7\times10^{-3}
    \left(\frac{|K|}{0.01}\right)
    \left(\frac{g_\mathrm{dec}}{200}\right)^{1/2}
    \left(\frac{M_{\tilde{g}}}{10^4\,\mathrm{GeV}}\right)^{-4}
    \left(\frac{H_I}{6\times10^{11}\,\mathrm{GeV}}\right)^2.
\end{align}
This condition is also consistent with the estimate of the baryon asymmetry
in Eq.~\eqref{eq:baryon_asymmetry_our_model}.

\subsection{The delayed-type Q-ball scenario}

In the delayed-type Q-ball scenario, the decay rate is almost independent of $\phi_\mathrm{osc}$, and hence $\alpha\simeq0$.
The cancellation of the correlated isocurvature perturbations requires
\begin{align}
    f_\mathrm{dec}
    \simeq
    0.471,
\end{align}
as follows from Eq.~\eqref{eq:compensation_fdec}.
The CMB constraint on the correlated isocurvature perturbations is again given by Eq.~\eqref{eq:corr_iso_const} and requires a fine-tuning of $f_\mathrm{dec}$ at the $1\%$ level.
The CMB constraint on the uncorrelated isocurvature perturbations, given in Eq.~\eqref{eq:uncor_obs_constr}, requires an additional fine-tuning of approximately $6\%$:
\begin{align}
    1.472
    <
    6\theta_\mathrm{osc}
    <
    1.670.
\end{align}
To reproduce the observed curvature perturbations, Eq.~\eqref{eq:phi_osc} requires the AD-field value to satisfy
\begin{align}
    \label{eq:phi_osc_Hubble_relation2}
    \phi_\mathrm{osc}
    \simeq
    1.1\times10^{15}\,\mathrm{GeV}
    \left(
        \frac{H_I}{10^{12}\,\mathrm{GeV}}
    \right).
\end{align}
The nonlinearity parameter estimated from Eq.~\eqref{eq:nonlinearity} is
\begin{align}
    f_\mathrm{NL}
    \simeq
    0.60,
\end{align}
which is consistent with the CMB constraint~\cite{Planck:2019kim}.

\begin{figure}[t]
  \centering
  \includegraphics[width=0.8\linewidth]{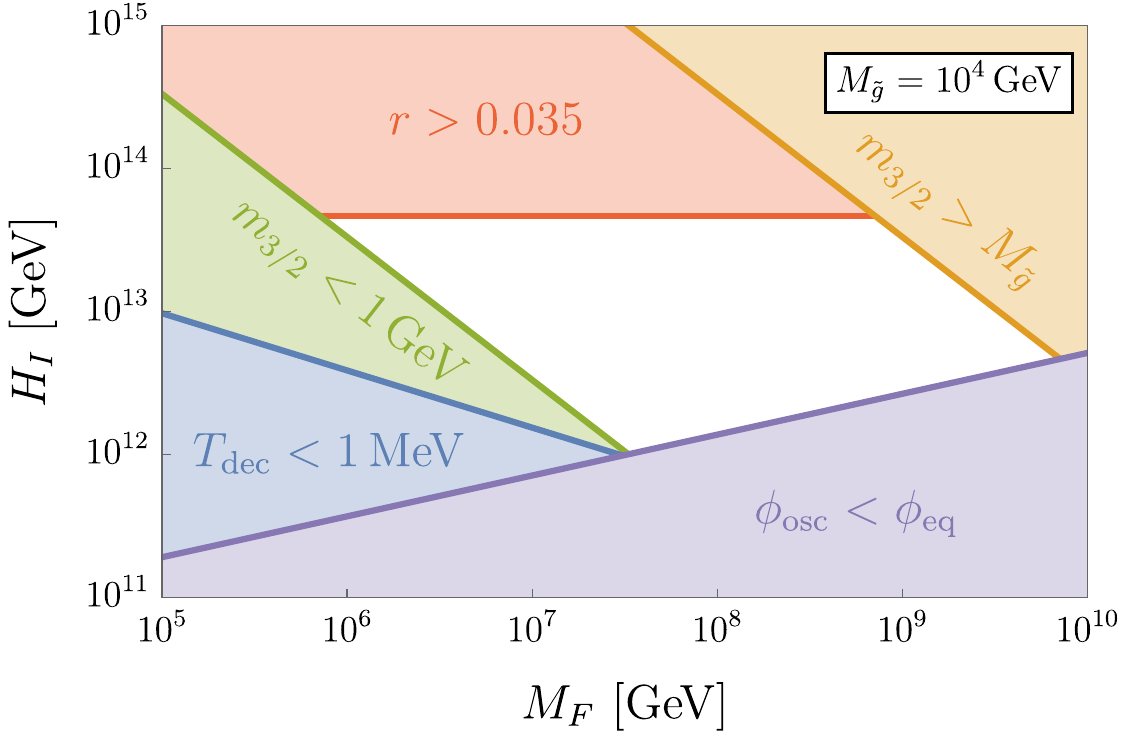}
  \caption{
  Constraints on the parameter space in the $(M_F,H_I)$ plane.
  The green region corresponds to $m_{3/2}<1\,\mathrm{GeV}$, for which Q-balls cannot decay into protons.
  The blue region is excluded by the BBN constraint, $T_{\mathrm{dec}}<1\,\mathrm{MeV}$.
  The red region is excluded by the observational bound on the tensor-to-scalar ratio, $r>0.035$.
  The purple region corresponds to $\phi_{\mathrm{osc}}<\phi_{\mathrm{eq}}$, where the assumed gravity-mediated-type Q-ball formation does not occur.
  The brown region is excluded because the gravitino is heavier than the gluino, $m_{3/2}>M_{\tilde g}$.
  The white region satisfies all the constraints.
  We take $M_{\tilde g}=10^4\,\mathrm{GeV}$.
  }
  \label{fig:constraint}
\end{figure}

Using Eqs.~\eqref{eq:phi_osc_Hubble_relation2}, \eqref{eq:decay_temp},
\eqref{eq:reheating_temp}, and \eqref{eq:gravitino}, we determine the model
parameters as follows:
\begin{align}
    T_\mathrm{dec}
    &\simeq
    5.1\,\mathrm{MeV}
    \left(\frac{g_\mathrm{dec}}{200}\right)^{1/6}
    \left(\frac{M_{\tilde{g}}}{10^4\,\mathrm{GeV}}\right)^{-10/3}
    \left(\frac{M_F}{10^8\,\mathrm{GeV}}\right)^{4/3}
    \left(\frac{H_I}{10^{12}\,\mathrm{GeV}}\right)^{10/3},
    \\[0.6em]
    T_R
    &\simeq
    1.5\times10^5\,\mathrm{GeV}
    \left(\frac{g_\mathrm{dec}}{200}\right)^{1/6}
    \left(\frac{M_{\tilde{g}}}{10^4\,\mathrm{GeV}}\right)^{-10/3}
    \left(\frac{M_F}{10^8\,\mathrm{GeV}}\right)^{4/3}
    \left(\frac{H_I}{10^{12}\,\mathrm{GeV}}\right)^{4/3},
    \\[0.6em]
    m_{3/2}
    &\simeq
    4.4\,\mathrm{GeV}
    \left(\frac{g_\mathrm{dec}}{200}\right)^{1/6}
    \left(\frac{M_{\tilde{g}}}{10^4\,\mathrm{GeV}}\right)^{-4/3}
    \left(\frac{M_F}{10^8\,\mathrm{GeV}}\right)^{4/3}
    \left(\frac{H_I}{10^{12}\,\mathrm{GeV}}\right)^{4/3}.
\end{align}
We show the allowed region in the $(M_F,H_I)$ parameter plane in Fig.\ref{fig:constraint}. 
Besides the constraints considered in the gravity-mediated-type Q-ball scenario, we also include the constraint $\phi_\mathrm{osc} > \phi_\mathrm{eq}$, which is given by
\begin{equation}
    H_I \gtrsim 1.4\times10^{12}\,\mathrm{GeV}\left(\frac{M_F}{10^8\,\mathrm{GeV}}\right)^\frac{2}{7}.
\end{equation}
As shown in the figure, the AD curvaton scenario is viable for 
\begin{align}
    10^6\,\mathrm{GeV}\,\lesssim  & ~M_F \lesssim \,10^{10}\,\mathrm{GeV},
    \\[0.4em]
    10^{12}\,\mathrm{GeV}\,\lesssim  & ~H_I \lesssim \,5\times 10^{14}\,\mathrm{GeV}.
\end{align} 

Finally, the condition under which the non-renormalizable term has a
negligible effect on the AD-field dynamics is given by
\begin{equation}
    \lambda
    \ll
    2.3\times10^{-6}
    \left(\frac{g_\mathrm{dec}}{200}\right)^{1/6}
    \left(\frac{M_{\tilde{g}}}{10^4\,\mathrm{GeV}}\right)^{-4/3}
    \left(\frac{M_F}{10^8\,\mathrm{GeV}}\right)^{4/3}
    \left(\frac{H_I}{3\times10^{12}\,\mathrm{GeV}}\right)^{-8/3}.
\end{equation}
This condition is again consistent with the estimate of the baryon asymmetry
in Eq.~\eqref{eq:baryon_asymmetry_our_model}.

\section{Summary and discussion}
\label{sec:summary}

We have revisited the AD curvaton scenario in the framework of SUSY, taking Q-ball formation into account, and have shown that the scenario can simultaneously explain the baryon asymmetry and the observed curvature perturbations of the Universe.
In this scenario, correlated baryon isocurvature perturbations are inevitably generated. However, their contribution to the correlated matter isocurvature perturbations can be canceled by the anti-correlated dark matter isocurvature perturbations associated with gravitino dark matter.
We have shown that a successful AD curvaton scenario can be realized for an appropriate choice of the Hubble parameter during inflation.

We have not specified a particular inflation model. 
Nevertheless, the inflation model must satisfy several conditions in the present scenario. 
Since we assume that the AD curvaton generates all of the observed curvature perturbations, the inflaton contribution to the curvature perturbations must be negligible. 
However, the inflationary dynamics still affect the spectral index of the curvature perturbation power spectrum.
In the AD curvaton scenario, assuming that the AD field is effectively massless during inflation, the spectral index is given by
\begin{equation}
    n_s
    =
    1-2\epsilon,
\end{equation}
where $\epsilon$ is the slow-roll parameter defined by
\begin{equation}
    \epsilon
    =
    \frac{M_{\mathrm{pl}}^2}{2}
    \left(
        \frac{V_I'}{V_I}
    \right)^2,
\end{equation}
with $V_I$ denoting the inflaton potential and $V_I'$ its derivative with respect to the inflaton field.
To reproduce the observed spectral index at the pivot scale $k_*$, $n_s=0.965\pm0.004$ at the $1\sigma$ level~\cite{Planck:2018jri}, the slow-roll parameter should lie in the range $\epsilon\simeq0.015\text{--}0.020$.
For example, a monomial chaotic inflation model with
$V_I\propto I^p$, where $I$ denotes the inflaton field, gives
\begin{equation}
    \epsilon
    \simeq
    \frac{p}{4N_*},
\end{equation}
where $N_*\simeq50\text{--}60$ is the number of $e$-folds corresponding to the pivot scale $k_*$. 
Therefore, a chaotic inflation model with a quartic potential, $p=4$, can provide a suitable inflationary background for the present scenario.

A relatively large value of $\epsilon$ generally points to large-field inflation models, many of which are excluded by the observational constraint on the tensor-to-scalar ratio, $r$, if the inflaton is responsible for generating the observed curvature perturbations. 
In the present scenario, however, the curvature perturbations are predominantly generated by the AD curvaton, thereby relaxing the constraint on the inflation model and allowing such large-field models to be consistent with observations. 
Furthermore, a hybrid scenario in which both the inflaton and the curvaton contribute to the curvature perturbations may further enlarge the viable parameter space~\cite{Fujita:2014iaa,Byrnes:2025kit}.

Since CMB observations cannot distinguish between baryon and dark matter isocurvature perturbations at the level of the linear power spectra, compensated isocurvature perturbations can evade the stringent CMB constraints on conventional matter isocurvature perturbations. 
However, 21cm observations are directly sensitive to baryon density fluctuations and can, in principle, distinguish between baryon and dark matter isocurvature perturbations~\cite{Kawasaki:2011ze,Hotinli:2021xln}. 
Therefore, future 21~cm observations may provide a distinctive test of the present scenario.

\begin{acknowledgments}
This work was supported by JSPS KAKENHI Grant Numbers JP25K07297 (M.K.) and JP25KJ1030 (S.N.), and by JST SPRING Grant Number JPMJSP2108 (S.N.). M.K. was also supported by the World Premier International Research Center Initiative (WPI), MEXT, Japan.
\end{acknowledgments}

\bibliographystyle{apsrev4-1}
\bibliography{Ref}

\end{document}